\documentclass[prl,twocolumn,superscriptaddress,amsmath,amssymb]{revtex4-1}
\usepackage{graphicx} % Include figure files
\usepackage{dcolumn}  % Align table columns on decimal point
\usepackage{bm}       % bold math
\usepackage{amsfonts}
\usepackage{color}
\newcommand{\blue}[1]{{\color{black} #1}}

\begin{document}

\title{Probing Electronic Fluxes via Time-Resolved X-ray Scattering}

\author{Gunter Hermann}
\affiliation{%
Institut f{\"u}r Chemie und Biochemie, Freie Universit{\"a}t Berlin, 
Takustra{\ss}e 3, D-14195 Berlin, Germany}
\affiliation{%
Department of Physics, Indian Institute of Technology Bombay,	
            Powai, Mumbai 400076  India}            

\author{Vincent Pohl}
\affiliation{%
Institut f{\"u}r Chemie und Biochemie, Freie Universit{\"a}t Berlin, 
Takustra{\ss}e 3, D-14195 Berlin, Germany}
\affiliation{%
Department of Physics, Indian Institute of Technology Bombay,
            Powai, Mumbai 400076  India}   
            
\author{Gopal Dixit}
\email[]{gdixit@phy.iitb.ac.in}
\affiliation{%
Department of Physics, Indian Institute of Technology Bombay,
            Powai, Mumbai 400076  India}

\author{Jean Christophe Tremblay}
\email[]{jean-christophe.tremblay@univ-lorraine.fr}
\affiliation{%
Institut f{\"u}r Chemie und Biochemie, Freie Universit{\"a}t Berlin, 
Takustra{\ss}e 3, D-14195 Berlin, Germany}
\affiliation{%
Laboratoire de Physique et Chimie Th\'eoriques, 
CNRS-Universit\'e de Lorraine, UMR 7019, ICPM, 1 Bd Arago, 57070 Metz, France}

\date{\today}

%\pacs{34.50.-s, 61.05.cf, 78.70.Ck}

%%%%%%%%%%%%%%%%% END OF PREAMBLE %%%%%%%%%%%%%%%%

\begin{abstract}
The current flux density is a vector field that can be used to describe theoretically how electrons flow in a system out-of-equilibrium.
In this work, we unequivocally demonstrate that the signal obtained from time-resolved X-ray scattering does not only map the time-evolution of the electronic charge distribution, but also encodes information about the associated electronic current flux density. We show how the electronic current flux density qualitatively maps the distribution of electronic momenta and reveals the underlying mechanism of ultrafast charge migration processes, while also providing quantitative information about the timescales of electronic coherences.
\end{abstract}

\maketitle
Time-resolved imaging of dynamically evolving electronic charge distribution is essential 
for complete understanding of complex chemical and biological processes in nature. 
Imaging of valence electron charge distribution is paramount to understand different instances during chemical reactions such as conformational changes, charge migration, and bond formation and breakage~\cite{lepine2014attosecond, leone2014will, remacle, lenz_jcp}. 
Following the quantum continuity equation, 
the flow of electron is accompanied  by associated fluxes~\cite{sakurai1967advanced}. 
The latter offers a wealth of  information and  has played a decisive role for understanding chemical reaction mechanisms~\cite{Barth2962,Barth7043,nagashima2009electron,takatsuka2011exploring,okuyama2012dynamical,diestler2013computation,takatsuka2014chemical,hermann2014electronic,yamamoto2015electron,hermann2016multidirectional, bredtmann2014x,okuyama2009electron,diestler2011coupled,patchkovskii2012electronic}. 
However, the notion of electronic fluxes has been restricted to theoretical modelling
\cite{barth2006unidirectional,okuyama2009electron,diestler2011coupled,patchkovskii2012electronic,kazuo2014chemical,pohl2016adiabatic,schild2016electronic,renziehausen2018many,schaupp2018time,matsuzaki2019electronic}
and there is no general way to probe  them directly in experiment.
In this work, we demonstrate theoretically real-space and real-time imaging of 
electronic fluxes associated with charge migration using time-resolved X-ray scattering (TRXS). 
For this purpose, we consider oriented benzene as a test system in which a pump pulse induces
adiabatic charge migration and ultrashort X-ray pulses probe the electronic fluxes accompanying charge migration. 

Scattering of X-rays from matter is an invaluable technique to unveil the
real-space structure of solids and molecules with atomic-scale resolution~\cite{Nielsen}.  
Tremendous technological progress has been made to generate 
tunable ultraintense and ultrashort pulses from X-ray free-electron lasers 
(XFELs)~\cite{ishikawa2012, pellegrini2016physics, emma2}. 
X-ray pulses with few femtoseconds pulse duration are routinely  generated at various XFELs (LCLS, SACLA, European XFEL).  
Moreover, few successful attempts have been demonstrated  
to generate attosecond X-ray pulses~\cite{tanaka2013, kumar2018generation, kumar2016temporally, shim2018isolated, hartmann2018attosecond,bucksbaum2019attoXR}. 
The availability of these ultrashort X-ray pulses offer to extend  X-ray scattering from static to time domain with 
unprecedented temporal resolution~\cite{lindroth2019challenges, young2018roadmap}.  
Scattering of ultrashort X-ray pulses from the evolving electronic charge distribution 
promises to provide stroboscopic snapshots of matter in action 
with atomic-scale spatial and temporal resolutions~\cite{bucksbaum2007, peplow2017next}.  
A direct approach to envision TRXS is a pump-probe experiment, where the pump pulse triggers the ultrafast dynamics and  
the induced dynamics is imaged by the ultrashort X-ray pulses.
Not only these ultrashort X-ray pulses allow to  map the motion of atoms in matter on the femtosecond timescale~\cite{peplow2017next, vrakking2016viewpoint},
but also to record movies of electronic motion taking place
from few femtoseconds to the attosecond timescale~\cite{dixit2012, vrakking2012}.  

The availability of ultrashort X-ray pulses has prompted 
TRXS experiments probing ultrafast processes with atomic-scale spatio-temporal resolutions.  
Static X-ray scattering from aligned 2,5-diiodobenzonitrile has been performed at LCLS~\cite{kupper2014x}. 
TRXS experiments allowed  imaging ultrafast vibrations in iodine~\cite{glownia2016self}.
Frequency-resolved TRXS was used to disentangle bound and dissociative electronic states during ultrafast vibrational dynamics in iodine~\cite{ware2019characterizing}. 
Photoinduced  structural change during ring opening electrocyclic chemical reaction in cyclohexadiene~\cite{minitti2015imaging, minitti2014toward}
and cis-trans photochemical structural changes in photoactive yellow protein~\cite{pande2016femtosecond} were captured by TRXS.  
Anisotropic TRXS measurements have been used to determine transition dipole moment and assign  excited electronics states in molecule~\cite{yong2018determining}.  
Different formalisms have been developed to simulate 
TRXS from non-equilibrium states of matter~\cite{cao1998ultrafast, henriksen, lorenz2010theory, dixit2012, dixit2013jcp, dixit2013prl, dixit2014theory, santra2014comment, 
bredtmann2014x, bennett2014time, dixit2017time}. 
It was demonstrated that the scattering signal obtained via TRXS from an electronic wavepacket is not associated with the Fourier transform of 
instantaneous electron density~\cite{cao1998ultrafast, henriksen, dixit2012, dixit2013jcp, bennett2014time}. 
Mukamel and co-workers have proposed that TRXS is capable to probe 
molecular nonadiabatic dynamics at avoided crossings and conical intersections~\cite{bennett2018monitoring, kowalewski2017monitoring}
Also, frequency- and wavevector-resolved TRXS has been used to probe the electron dynamics in molecules~\cite{bennett2014time}. 
Recently, it was shown that TRXS can probe electronic coherences among electronic states~\cite{simmermacher2019electronic},
and that TRXS from diatomic molecules are not centrosymmetric~\cite{starace2019pra}.

The main focus of this work is to illustrate the capability of TRXS for imaging quantum 
fluxes during non-stationary charge migration in a coherent electronic wavepacket prepared by an ultrashort pump pulse. 
Quantum fluxes find their origin in the interferences among quantum mechanical phases.
The time-resolved response signal that can be extracted from a TRXS experiment
contains information about these electronic coherences, and it is therefore suitable for mapping the current  flux density.
By following the time-evolution of the coherent electronic wavepacket, we demonstrate the relation between the quantum continuity equation for non-stationary charge migration and TRXS. 

\begin{figure}[htb!]
\includegraphics[width=\linewidth]{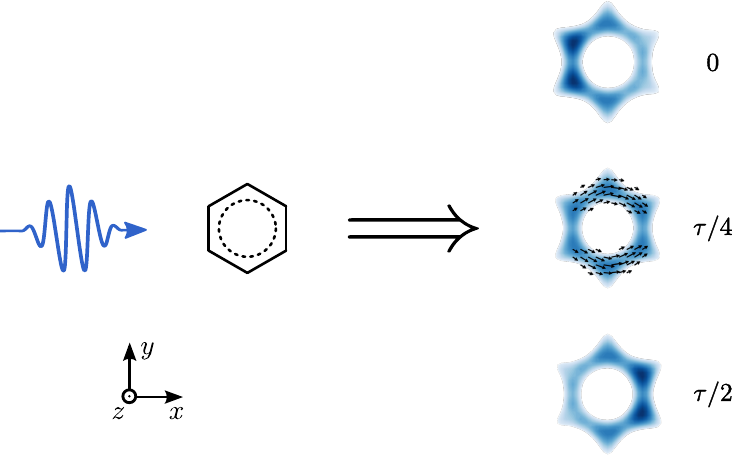}
\caption{Conceptual sketch of the charge migration mechanism. An $x$-polarized pump pulse induces non-stationary charge migration associated with an electronic wavepacket in benzene. The electron density (blue shaded area) migrates from one side of the molecule to the other with a period $\tau = 504$ attoseconds. Black arrows correspond to the electronic flux density associated with this process (for $z=1\,{\rm a_0}$).} \label{fig01}
\end{figure} 
In this work, we investigate charge migration in benzene induced by a linearly $x$-polarized pulse. 
A resonant pump pulse of 3.57~fs duration (92~meV bandwidth) and 0.6 TW/cm$^{2}$ intensity at 8.2 eV photon energy
is used to prepare a coherent electronic superposition of the $A_{1g}$ ground state
and a low-lying optically accessible $E_{1u,x}$ electronic states~\cite{jia2017quantum}.
The time period of the non-stationary charge migration corresponds to $\tau = $504 attoseconds (see Fig.~\ref{fig01}).
The timescale of the electronic motion of the wavepacket is much faster than the motion of nuclei \cite{mineo2014vibrational,despre2015attosecond},
which are kept frozen. 
The state-averaged CASSCF(6,6) method implemented in MOLPRO~\cite{werner2012molpro} is used
with an aug-cc-pVTZ basis~\cite{dunning1989gaussian} to compute
the singlet ground and low-lying electronic excited states of benzene, which is aligned in the $xy$-plane.  
As in Ref.[\citenum{hermann2016multidirectional}], Multi-Reference Configuration Interaction with Single and Double excitations is employed to correct excitation energies. 

\begin{figure*}[htb!]
\includegraphics[width=\textwidth]{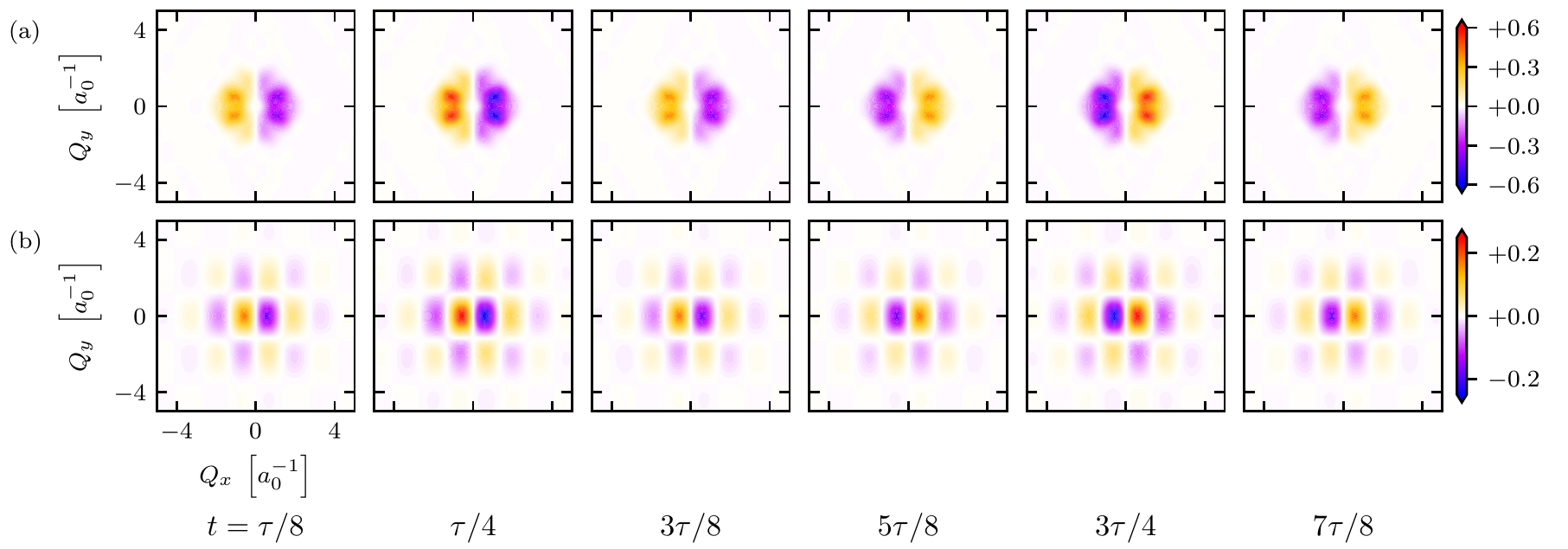}
\caption{Time-resolved signals in the $Q_{x}-Q_{y}$ plane ($Q_{z}= 0$) as a function
of pump-probe delay times with a  charge migration period of $\tau = 504$ attoseconds for benzene. 
The time-resolved signal at time zero $t = 0$ is subtracted from the signals at later delay times. 
(a) Scattering patterns obtained using Eq.~(\ref{eq1}), and
(b) time-derivative of the momentum-space electron density $\rho(\mathbf{Q})$. 
The intensity of the scattering patterns are presented in units of ${dP_{e}}/{d\Omega}$.} \label{fig1}
\end{figure*} 
To image charge migration and the associated fluxes in benzene,  
the time-resolved scattering signal is simulated using an expression for the differential scattering probability (DSP)
of the form (in atomic units)~\cite{henriksen, dixit2012, dixit2014theory} 
 \begin{equation}\label{eq1}
\frac{dP}{d\Omega} =
\frac{dP_{e}}{d\Omega} \sum_{j} \left| 
\int d\mathbf{r} ~\langle \psi_{j}  | \hat{n}(\mathbf{r}) | \Phi({t}) \rangle~ e^{-i \mathbf{Q} \cdot \mathbf{r}}\right|^{2}, 
\end{equation}
where $\frac{dP_{e}}{d\Omega}$ is the Thomson scattering cross section of a free electron, 
$| \psi_{j}  \rangle$ is an eigenstate of the unperturbed Hamiltonian, 
$| \Phi({t}) \rangle$ is an electronic wavepacket with $t$ as pump-probe time delay, 
$\hat{n}(\mathbf{r})$ is the electron density operator, and
$\mathbf{Q}$ is the photon momentum transfer.  
In previous work, the numerical simulation of TRXS from the electronic wavepacket 
has been limited to  atomic and simple molecular systems~\cite{dixit2012, dixit2013jcp, bennett2014time, simmermacher2017time, simmermacher2019electronic}.
\blue{For a general electronic wavepacket, the summation over $j$ in Eq.~({\ref{eq1}}) runs over a complete set of eigenstates.
Simulating scattering signals using a large number of eigenstates is usually not 
practical due to the associated computational cost. 
The scattering  signal is shown to converge rapidly with respect to the number of 
eigenstates (see Fig. S2).  
All results reported in this work are computed using the 7 lowest-lying of eigenstates. 
All transition amplitudes of the density operator from the many-body eigenfunctions, i.e.,
$\langle \psi_{A_{1g}}  | \hat{n}(\mathbf{r}) | \psi_j \rangle$ and
$\langle \psi_{E_{1u,x}}  | \hat{n}(\mathbf{r}) | \psi_j \rangle$, are simulated using 
the ORBKIT toolbox~\cite{hermann2016orbkit, pohl2017open, hermann2017open}.
In the past, the summation over $j$ was restricted to the  eigenstates spanning the 
wavepacket and the simulated scattering signals were used to understand the measured 
signals~\cite{glownia2016self, ware2019characterizing, yong2018determining, minitti2015imaging, minitti2014toward}.    Historically, it was believed that the DSP 
is proportional to the instantaneous electron density of the wavepacket.  Neglecting the effect of electronic coherences was shown to be  incorrect in similar contexts~\cite{cao1998ultrafast, henriksen, dixit2012, dixit2013jcp, bennett2014time}.} 

Here, we observe  that the  time-dependence of the momentum-space density also
differs from that of the current flux density.
The time-evolution of the signal obtained from Eq.~\eqref{eq1} correlates with the time-derivative of the density calculated from first principles.
The theoretical support for this correspondence is detailed in the Supporting Information (SI), where the time-evolution of the signals is derived for a general superposition state.
To confirm these results numerically, we investigate the many-electron dynamics using the instantaneous variation of the one-electron density, $\partial_t\rho(\mathbf{r},t)$, and the associate current flux density, $\mathbf{j}(\mathbf{r},t)$.
These are connected via the continuity equation, $\partial_t\rho(\mathbf{r},t) = -\vec{\nabla}\cdot\mathbf{j}(\mathbf{r},t)$.
The one-electron density and the current flux density are computed from the time-dependent many-electron wavepacket, as  described elsewhere \cite{pohl2017open, hermann2017open}.

Time-resolved scattering patterns corresponding to an electronic wavepacket for different pump-probe delay times are presented
in Fig.~\ref{fig1}a. The electronic wavepacket consists of a coherent superposition of 
two many-body electronic  states which evolves according to
\begin{equation}\label{wp}
\Phi(\mathbf{r}^N,t) = c_{A_{1g}}(t)\psi_{A_{1g}}(\mathbf{r}^N) + c_{E_{1u,x}}(t)\psi_{E_{1u,x}}(\mathbf{r}^N)
\end{equation}
The coefficients $c_j(t)= 2^{-1/2}e^{-i \varepsilon_jt/\hbar}$ are associated with the ground state,
$\psi_{A_{1g}}(\mathbf{r}^N)$ at energy $\varepsilon_{A_{1g}}$,
and an optically accessible excited state, $\psi_{E_{1u,x}}(\mathbf{r}^N)$ at energy $\varepsilon_{E_{1u,x}}$.
Eq.~(\ref{eq1}) is used to simulate the  patterns shown in Fig.~\ref{fig1}a and presented in the $Q_{x}-Q_{y}$ plane ($Q_{z}= 0$). 
For representation purposes, the scattering pattern at $t = 0$ is subtracted. 
The scattering patterns at $t = \tau/4$ and $3\tau/4$ have opposite phase,
whereas they are similar at $t = \tau/8$ and $3\tau/8$, and at $ t = 5\tau/8$ and $7\tau/8$.
Hence, the scattering patterns are sensitive to delay times, with a $\sin$ behaviour of period $\tau$. 
The time-derivative of the momentum space electron density,
$\partial_t\rho({\mathbf{Q}})$, is shown in the central panels of Fig.~\ref{fig1}b.
As visible from  Figs.~\ref{fig1}a and ~\ref{fig1}b, there is a one-to-one correspondence between the time-evolution of the scattering patterns
obtained from Eq.~\eqref{eq1} and the time-derivative of the electron density.
Although the structure of $\partial_t\rho({\mathbf{Q}})$ extends further in the $Q_{x}-Q_{y}$ plane, it contains the information of the DSP signal
and the two quantities have the same period.
As shown in the SI, the DSP signal is not exactly the time-derivative of the momentum-space electron density,
but rather the convolution of its different contributions. This is the reason why the timescales correlate exactly with the dynamics but the spatial extent is different in both cases.
According to Eq.\,(S12), the DSP signal from Eq.\,\eqref{eq1} also contains a 
contribution from the instantaneous density with $\sin^2$ dependency of period $2\tau$, which would lead to an asymmetry in the signal at
$t=\tau/4$ and $3\tau/4$. This asymmetry is not observed since the associated term is
vanishingly small (see Fig.\,S1). Hence, the time-evolution of the experimental DSP signal yields
quantitative information about the timescales involved in the {\it time-derivative} of the one-electron density.
\begin{figure}[htb!]
\includegraphics[width=0.9\linewidth]{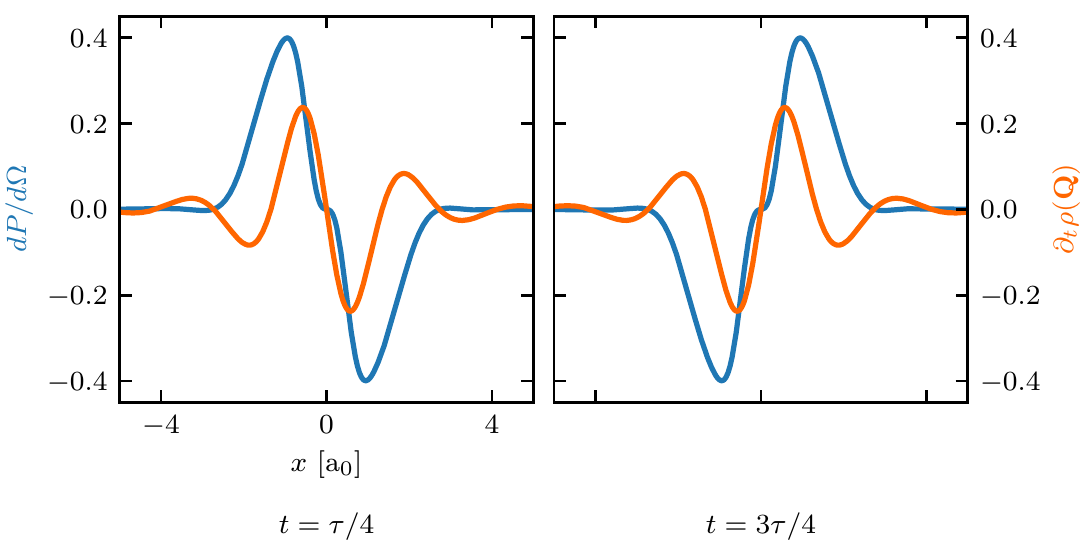}
\caption{Comparison of the time-resolved signals from Eq.\eqref{eq1} (blue) and the time-derivative of the momentum-space electron density $\rho(\mathbf{Q})$ (orange)
along $Q_{x}$ ($Q_{y}=Q_{z}= 0$) at pump-probe delay times $t = \tau/4$ and $t = 3\tau/4$.} \label{fig1-1}
\end{figure} 
A more quantitative comparison can be obtained from Fig.\,\ref{fig1-1}, which shows 1D cut of the DSP and 
the time-derivative of the momentum-space electron density. Despite differences at higher momenta,
the pictures that emerges at low momenta are in good agreement.
Low momenta are the most important, as they were shown to map the dynamics of valence electrons \cite{bredtmann2014x}.

%On the other hand, the scattering pattern obtained using Eq.~\eqref{eq2} does not correspond with the quantum continuity equation (see Fig.~\ref{fig1}c).
%The patterns at $t = \tau/8,  3\tau/8,  5\tau/8$ and $ 7\tau/8$ are identical and the sign of the DSP difference signal remains positive at all times.
%Further, the patterns at $t = \tau/4$ and $3\tau/4$ are the same, whereas the charge migration at these two times are complete opposite (see Fig.~\ref{fig01}). In a two-level system, the expression for the signal
%Eq.\,\eqref{eq2} exhibits approximately a $\cos^2$ behaviour of period $\tau$ (c.f. SI).
%This confirms that the scattering patterns obtained from Eq.~\eqref{eq2} predict the wrong
%time-dependency for the electronic motion (c.f.~Fig.~\ref{fig1}c), and that the Fourier transform
%of the instantaneous electron density is not adequate to describe non-stationary charge migration.

\begin{figure*}[ht!]
\includegraphics[width=\textwidth]{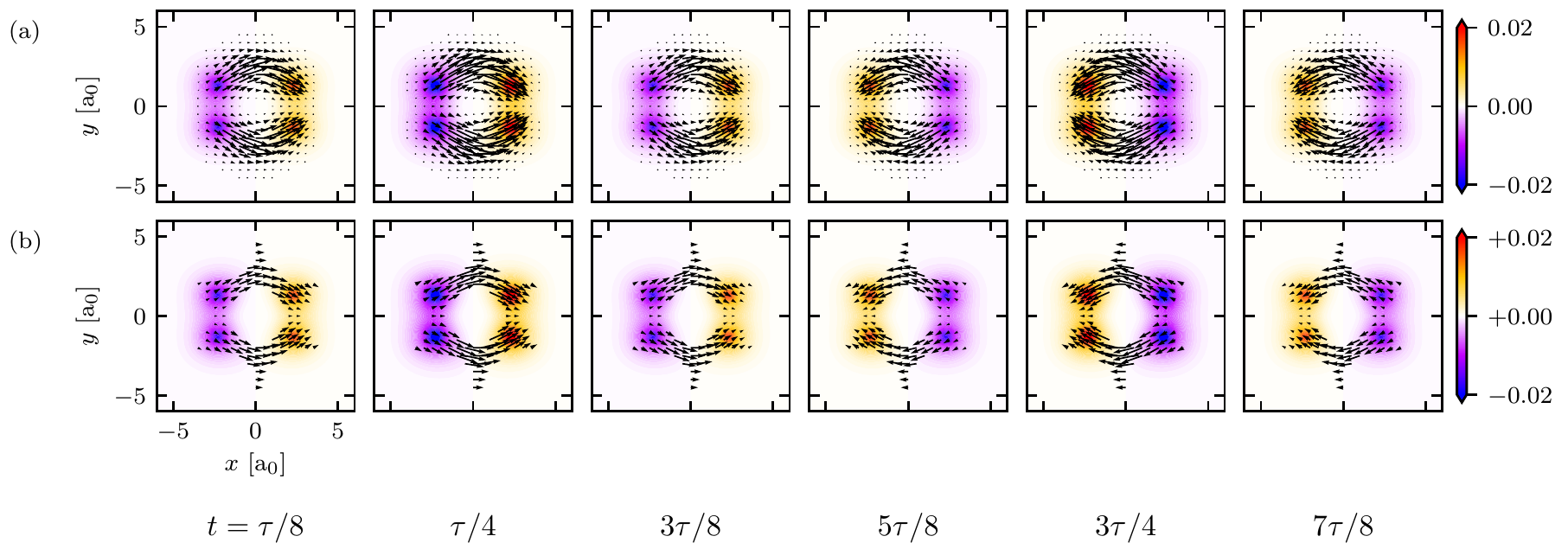}
\caption{Panel a: Fourier transform of the time-derivative of the momentum-space electron density $\rho(\mathbf{Q})$ 
(colour map) and the associated current flux density (arrows)  as a function of pump-probe delay times.
Panel b: Fourier transform of the time-derivative  of the momentum-space electron density $\rho(\mathbf{Q})$ 
(colour map) and velocity field, $\mathbf{v}(\mathbf{r},t)=\mathbf{j}(\mathbf{r},t)/\rho(\mathbf{r},t)$, (arrows) 
depicted only at positions where the electron density is above $\rho(r) > 10^{-10}\mathrm{a}_0^{-3}$.} \label{fig2}
\end{figure*}

The time-dependent DSP signal encodes information about the time-evolution of the wavepacket in momentum-space.
Hence, it contains information related to the velocity distributions.
To reveal this information, we first reconstruct the current flux density from the many-electron wavepacket associated with the charge migration in benzene. 
The current flux density is a vector field in configuration space that maps the displacement of the volume elements of the one-electron density.
Fig.~\ref{fig2} presents the time-derivative of the electron density (colour map) and the current flux density (arrows) at various delay times. 
These quantities are related via the electronic continuity equation, which describes the many-electron dynamics as
the flow of a strongly correlated electronic fluid.
The one-electron density is seen to migrate from left (violet/blue) to right (yellow/red) in the first half period of the charge migration process,
before coming back.
The nodal plane along the $y$-axis, which is a consequence of the pump pulse used to generate this superposition
state \cite{jia2017quantum}, is retained at all times.

The mechanistic information of the charge migration is encoded in the scattering patterns.
However, it is not easy to know where are localised electrons that move in a certain region directly from the patterns. 
As can be seen from Fig.~\ref{fig2}a, the direction of the arrows correlates qualitatively with the time-resolved scattering
patterns in momentum space depicted in the upper panels of the previous figure (see Fig.~\ref{fig1}a). 
The dominant electron flow is along the $x$-direction, with minor components in the $Q_{x}-Q_{y}$ plane at angles corresponding to the C-C bonds
of the molecular scaffold.
Both pictures are consistent and describe a bond-to-bond electron migration mechanism.
On the other hand, the time-derivative of the one-electron density reveals a more intricate nodal structure in the central panels of Fig.~\ref{fig1}b.
As discussed in previous work \cite{pohl2017open, hermann2017open},
the derivative of the electronic density around the nuclei is sensitive to the choice of atomic basis set.
The Fourier transform of the density reveals this sensitivity in momentum space.

The velocity field, calculated as $\mathbf{v}(\mathbf{r},t)=\mathbf{j}(\mathbf{r},t)/\rho(\mathbf{r},t)$, offers an alternative representation
of the charge migration mechanism.
It is shown in Fig.~\ref{fig2}b (arrows), along with the time-derivative of electron density (colour map). 
%The velocity field is only depicted at positions where the electron density is above a certain threshold.  
Although it contains mostly the same information as the current flux density, the velocity field is more easily related to the momentum
observed in the DSP signal. The time-dependent rescaling through the one-electron density yields a better contrast of the electronic flow,
which simplifies the direct comparison with scattering patterns. It can be observed that the electrons flow faster around the central carbon atoms,
which contrasts with the picture offered by the current flux density. The latter predicts an homogeneous $\pi$-electron flow
along the two C-C-C units of the scaffold. The $\pi$-electron density is lower on the atoms than on the bonds. 
Rescaling the flux density by the density thus reveals an increased velocity at the central carbon atoms. This phenomenon
is analogous to the Venturi effect in classical hydrodynamics, if we assimilate the reduction of the electron density to a reduction
of the cross-section through which electrons flow. Since the volumetric flow rate is conserved,
the smaller electron density implies an increased velocity and a reduced hydrodynamic pressure at the carbon atoms.

In conclusion, we have shown that ultrafast time-resolved X-ray scattering has potential to extract mechanistic information about the flow of electrons
in a molecule out-of-equilibrium by mapping the electronic current flux density. The latter is related to the time-variation of the momentum-space density.
The TRXS signal contains qualitative information about the instantaneous electronic velocity distribution
and quantitative information about temporal electronic coherences.
Cross-correlation with first-principle simulations can be used to reveal the electronic flux density,
which contains the time- and space-resolved mechanistic details of the electron migration process.
The experimental realization is limited by the time- and momentum-resolutions of TRXS. While benzene is beyond 
current experiments, the prediction remains valid for slower processes.
This would require including nuclear motion in the theoretical treatment.

\section*{Acknowledgements}
G.D. acknowledges for the Ramanujan fellowship (SB/S2/ RJN-152/2015). 
G.H. and V.P. are grateful for travel funding of the Freie Universit\"at Berlin through the "Indo-German Partnership in Higher Education"
program of the DAAD.
J.C.T. , G.H., and V.P. are thankful to the Deutsche Forschungsgemeinschaft for funding through grant TR1109/2-1.

%\bibliography{imaging} 

\end{document}